# Collaborative Boundary-aware Context Encoding Networks for Error Map Prediction

Zhenxi Zhang[1], Chunna Tian[1], Jie Li[1], Zhusi Zhong[1], Zhicheng Jiao[2], and Xinbo Gao[1]

[1] School of Electronic Engineering, Xidian University, Xi'an 710071, China
zxzhang_5@stu.xidian.edu.cn
[2] Perelman School of Medicine at University of Pennsylvania

**Abstract.** Medical image segmentation is usually regarded as one of the most important intermediate steps in clinical situations and medical imaging research. Thus, accurately assessing the segmentation quality of the automatically generated predictions is essential for guaranteeing the reliability of the results of the computer-assisted diagnosis (CAD). Many researchers apply neural networks to train segmentation quality regression models to estimate the segmentation quality of a new data cohort without labeled ground truth. Recently, a novel idea is proposed that transforming the segmentation quality assessment (SQA) problem into the pixel-wise error map prediction task in the form of segmentation. However, the simple application of vanilla segmentation structures in medical image fails to detect some small and thin error regions of the auto-generated masks with complex anatomical structures. In this paper, we propose collaborative boundary-aware context encoding networks called AEP-Net for error prediction task. Specifically, we propose a collaborative feature transformation branch for better feature fusion between images and masks, and precise localization of error regions. Further, we propose a context encoding module to utilize the global predictor from the error map to enhance the feature representation and regularize the networks. We perform experiments on IBSR v2.0 dataset and ACDC dataset. The AEP-Net achieves an average DSC of 0.8358, 0.8164 for error prediction task, and shows a high Pearson correlation coefficient of 0.9873 between the actual segmentation accuracy and the predicted accuracy inferred from the predicted error map on IBSR v2.0 dataset, which verifies the efficacy of our AEP-Net.

## 1 Introduction

Image segmentation plays a fundamental role in medical imaging research and clinical situations such as radiotherapy and image-guided interventions. In recent years, deep learning methods have achieved improvements in various medical image segmentation tasks. However, it is challenging to obtain promising segmentation results when lacking enough manual annotations. Thus, it is of vital importance to employ an accurate segmentation quality assessment of per case for reducing errors in subsequent analysis or downstream procedures, and it is also helpful to the clinicians. However, it's unrealistic to get all manually labeled ground truths (GTs) of testing scans. Therefore, many efforts have been taken to SQA methods without ground truth. The reverse classification accuracy (RCA) method [1] trained a reverse classifier with the generated mask, and utilized the segmentation performance on a set of reference images as the proxy assessment. The main drawback of this method is the high cost of computational time. Some methods [2,3] trained the regression networks to predict non-reference quality metrics. For example, [2] regressed the similarity between a degraded vessel tree and a manual segmentation. [3] directly regressed the Dice Similarity Coefficient (DSC) of the generated masks. These methods need a large number of training samples, and they are vulnerable to the adversarial attacks. The uncertainty measures between multiple candidate segmentation masks have become another popular method to assess segmentation quality automatically, such as variation or DSC agreements between Monte Carlo samples in [4]. However, they usually need multiple forward propagations of models to get candidate masks, increasing the computational overhead and reducing efficiency.

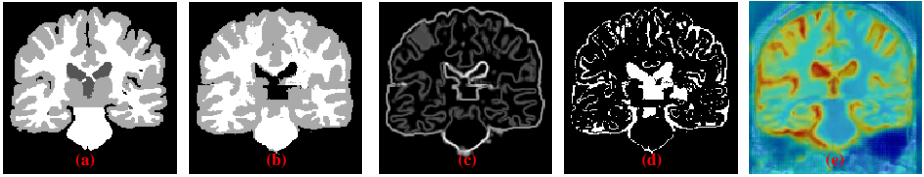

**Fig. 1.** (a) Ground truth (b) Segmentation mask (c) Enhanced class boundary (d) Error map: the misclassified pixels are set to class: 0 (white region), while the correct pixels are set to class: 1 (black region) (e)The attention map from the boundary-aware attention module.

### 1.1 Error Map Prediction

Recently, Zhang R, et al. [5] proposed an error map prediction network to predict the misclassified pixels in the generated masks, which transferred the quality assessment task into a segmentation task. The high potential of this method is that it can provide clinicians with a visual guide by pointing out the mislabeled pixels, and the quality metric also can be inferred from the predicted error map. However, this task is different from the general segmentation frameworks. It cannot achieve satisfactory results on the dataset which has complex anatomical structures when just using the vanilla structure, such as 3D U-Net [9], VoxResNet [10], and directly concatenating the original images with masks as network input. Specifically, there is a big semantic gap [6] between the original images and the segmentation masks. The low-level features from the original images are too noisy to provide more useful information. Secondly, the mislabeled pixels are prone to appear on the semantic boundary, the vanilla structure always fails to highlight the information of semantic boundary. Last but not least, the error map itself persists the global contextual supervision information representing segmentation quality. Understanding and utilizing this information is essential for this task.

**Contribution**: To address these problems, we propose an accurate error prediction framework called AEP-Net, which consists of three parts: the main error-prediction (MEP) branch, the collaborative boundary-aware feature transformation (CBFT) branch, the context encoding unit (CEU) [7]. Specifically, the CBFT integrates class boundary information to enhance the feature fusion and provide guidance for error detection. In addition, the CEU imports the error map predicator as the global information constraint to boost the feature representation of the networks. Under the cooperation of these three parts, we achieve higher average DSCs for error map segmentation on IBSR v2.0 and ACDC datasets, and observe a strong positive correlation between the predicted segmentation accuracy and the actual segmentation accuracy.

## 2 Method

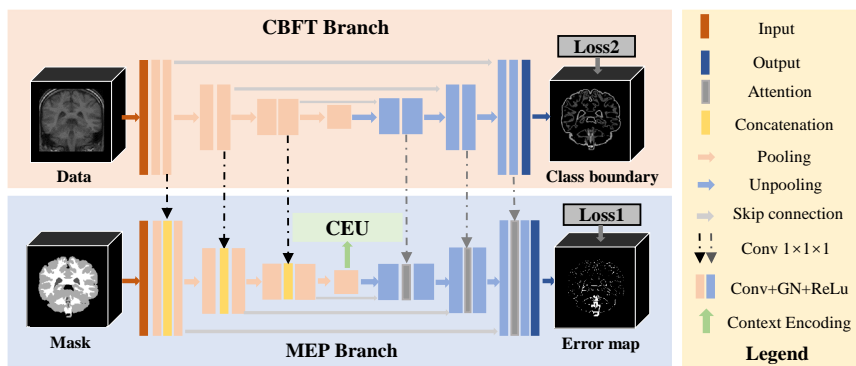

**Fig. 2.** The proposed framework, which consists of the MEP branch, the CBFT branch and the CEU module.

### 2.1 Overview

Fig. 2 illustrates the architecture of our AEP-Net, which is based on the MEP branch, the CBFT branch and the CEU. We leverage the idea of feature sharing [8] and the attention mechanism to build the collaborative relationships between the MEP and the CBFT. Maintaining the class boundary in the CBFT can capture more useful features near the class boundaries, which are also error-prone regions. Thus, this collaborative branch can provide more valuable transformed features and fine-grained attention for the MEP branch. The MEP takes the one-hot masks as input, and it obtains the upgraded features from the CBFT branch during the forward process. Both two branches are based on encoder-decoder structures. Additionally, the global quality information derived from the actual error map is utilized to regularize the networks through the CEU.

## 2.2 Main Error Prediction Branch

We employ a u-net structure to predict error map. The two-class Generalized Dice Loss [12] is used as the error map segmentation loss, since it performs better for class imbalance problem. Let $R$ be the reference error map (one-hot ground truth) with values $r_i(c)$ for class c, and $P$ be the predicted two-channel error map, $p_i(c)$ be the predicted probability of voxel $i$ for class $c$. It can be formulated as:

$$loss_1(P, R) = 1 - 2 \frac{\sum_{c=0}^{1} \omega_c \sum_i p_i(c) r_i(c)}{\sum_{c=0}^{1} \omega_c \sum_i p_i(c) + r_i(c)} \quad (1)$$

where $\omega_c = 1/(\sum_{i=1}^{N} r_i(c))^2$. For better feature fusion of the two branches at different semantic levels, we employ the different feature sharing methods. Specifically, at the encoder stage, the relatively low-level local features from the CBFT branch are concatenated with the features from the MEP branch. While at the decoder stage, we boost the feature fusion by introducing the boundary-aware attention module. Let the $F_{mep}$ and the $F_{cbft}$ denote the features from the MEP branch and the CBFT branch respectively, the $C_{1\times1\times1}$ denote the 1×1×1 convolution, the process of attention module is described in Eq.2.

$$F_{out} = F_{mep} + F_{mep} \otimes \sigma\left(C_{1\times1\times1}(F_{cbft})\right) \quad (2)$$

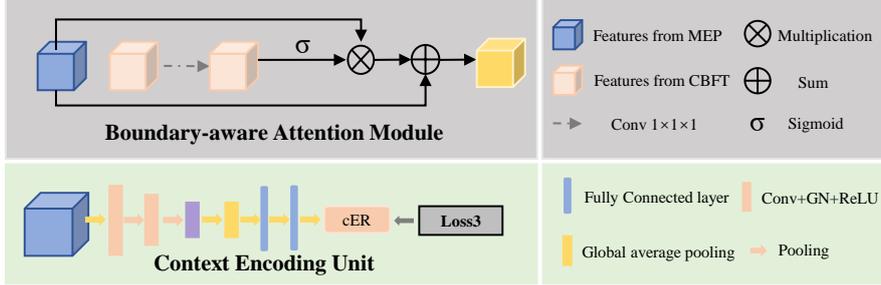

**Fig. 3.** Boundary-aware Attention Module and Context Encoding Unit

## 2.3 Boundary-aware Feature Transformation Branch

As stated in section 1.1, there is a giant semantic gap between the original data and the segmentation result. The direct fusion of high-level segmentation result with the pure low-level original data helps little. Thus, we propose a soft-boundary detection network to transform the features of the original data gradually. We also employ a u-net which keeps pace with the MEP branch. The enhanced boundary, which reflects the semantic class changes between different tissues and reveals the error-prone region, is generated by applying a 3D Sobel layer to the generated mask $Y$ and a simple boundary enhancement method. Assume that the total number of voxels is $N$, let $X$ denote the original data, $\varphi(\cdot)$ represent the network parameters of the CBFT branch, $S$ denote the Sobel gradient of $Y$. In all positions where the gradient is greater than 0, we add a constant of the max gradient of $S$, which ensures the values of the predicted boundary are all greater than 0.5 after the normalization. Thus, the enhanced boundary $B$ can be formulated as:

$$b_i = \begin{cases} (s_i + \max_{i=1,2...,N} s_i)/(2 \times \max_{i=1,2...,N} s_i), & if\ s_i > 0 \\ 0, & otherwise \end{cases} \quad (3)$$

The MSE loss is used in the CBFT module. Thus, the $loss_1$ can be defined as:

$$loss_2(X, B) = \frac{1}{N} \sum_{i=1}^{N} (b_i - \varphi(x_i))^2 \quad (4)$$

This branch has two major benefits. Firstly, it guarantees that the shared features persist progressively enhanced semantic information. Secondly, it promotes the MEP to focus on the error-prone semantic boundary by providing attention maps as shown in Fig. 1 (e).

### 2.4 Context Encoding Unit

In order to exploit the global contextual information from the real error map, we define the real error rate ($rER$), which is inferred from the real error map, as the proportion of the misclassified voxels to the total voxels. Then we employ a Context Encoding Unit (CEU) to enhance the feature representation of the MEP by regressing the $RER$. The CEU is attached to the end encoder layer of the MEP. The CEU begins with 3×3×3 convolution layers with group normalization and ReLU, and reduces the spatial dimension by the 3D max-pooling, and the global average pooling operations, ends with two fully connected layers and a sigmoid activation function. We define the error rate predicted by the CEU as $cER$. The loss of this module can be defined as: $loss_3 = (cER - rER)^2$. Thus, the total loss of our proposed AEP-Net is defined as: $loss_{total} = \alpha \cdot loss_1 + \beta \cdot loss_2 + \gamma \cdot loss_3$, where $\alpha$, $\beta$, $\gamma$ are set as 0.3, 0.3, and 0.6 via parameters selection experiment.

## 3 Experiments and Results

### 3.1 Datasets and Mask Generation

The AEP-Net is evaluated on two datasets. One is the brain anatomical segmentation dataset IBSR v2.0 (https://www.nitrc.org/projects/ibsr), which has 18 T1-weighted MR images of 256 ×128×256 voxels, and manually labeled segmentation for three tissues, white matter, grey matter, cerebrospinal fluid. Another one is the training set of the Automated Cardiac Diagnosis Challenge (ACDC) dataset [11], which consists of 100 4D short-axis cine-CMR scans, and has manual reference images of the right ventricular cavity (RV), the myocardium (Myo), and the left ventricular cavity (LV) in ED (end-diastole) and ES (end-systole) phases, respectively. The in-plane dimension of all scans has a wide range, from 154×224 to 428×512.

The candidate segmentation masks are generated based on 3D U-Net and VoxResNet, two classical segmentation models for volumetric medical images. In IBSR v2.0 dataset, we implement three 3D U-Nets with different number of initial feature maps: 16, 24, and 32, and other operations are same. Then we collect segmentation masks from three training epochs. For VoxResNet, one side output by the deep supervision loss and the final output are generated as two segmentation masks. Further, we also collect masks from two training epochs. Thus, each scan has 13 generated masks. In ACDC dataset, we collect four groups of 800 masks from two 3D U-Nets and VoxResNet. The weighted Dice score histograms of two datasets are shown in supplementary materials. Then we produce the actual error maps by comparing the generated masks with the manual ground truths.

### 3.2    Implementation details

The evaluation is implemented with three-fold cross validation on both two datasets. The generated masks and the real error maps of the training sets are further used to train the AEP-Net. For mask generation, we utilize the cross-entropy loss and multi-class Dice loss to train the segmentation networks in a standard manner. For the error prediction task, we utilize the proposed $loss_{total}$ to train the AEP-Net in an end-to-end way. For data processing and augmentation, we apply the z-score standardization and normalize the voxel intensity to [0,1] on all scans. During the training, we use a random crop size 128×32×128 for brain images, and a size 200×200×8 for cardiac images. Since the dimensions of some scans in ACDC are smaller than this crop size, we utilize zero-padding to make these scans meet the minimum input requirements. Further, we apply the random mirror flipping (along $x, z$ axes for brain and $x, y$ axes for cardiac). We implemented our networks on PyTorch and trained them on a single NVIDIA 1080 Ti GPU with 11GB RAM. The Adam optimizer with the poly learning rate policy was used in our experiments, where the initial learning rate is $1e^{-3}$, the poly power is 0.9. It took about 36 hours to train the AEP-Net on both two datasets. The time cost of predicting error map by forward propagation is 14 seconds for one brain MRI scan, and 3.2 seconds for one cardiac scan on average.

### 3.3    Evaluation Metrics

To evaluate the efficacy of the proposed AEP-Net, we explain the indicators we use as follows. The $Seg.DSC$ and $Seg.Acc$ means the dice similarity coefficient and accuracy between the generated masks and the ground truths, while the $DSC$, $Acc$, $Precision$ ($Prec$), and $Recall(Recl)$ are calculated between the real error map and the predicted error map.

**Table 1.** The error prediction results of the IBSR v2.0 and ACDC datasets. The $(x, y]$ denotes the $DSC$ range.

| Mask type | Num | Seg.DSC | Seg.Acc | DSC | Acc | Prec | Recl |
|---|---|---|---|---|---|---|---|
| (0.5,0.6] | 36 | 0.5696 | 0.9125 | 0.8982 | 0.9852 | 0.8539 | 0.9489 |
| (0.6,0.7] | 45 | 0.6457 | 0.9193 | 0.9074 | 0.9868 | 0.8894 | 0.9281 |
| (0.7,0.8] | 46 | 0.7600 | 0.9457 | 0.8900 | 0.9897 | 0.9113 | 0.8702 |
| (0.8,0.9] | 76 | 0.8495 | 0.9654 | 0.8119 | 0.9897 | 0.8211 | 0.8152 |
| (0.9,0.95] | 31 | 0.9215 | 0.9787 | 0.6376 | 0.9879 | 0.5341 | 0.7964 |
| IBSR v2.0 | 234 | 0.7592 | 0.9463 | **0.8358** | **0.9882** | **0.8190** | **0.8658** |
| ACDC | 800 | 0.5782 | 0.8782 | 0.8164 | 0.9880 | 0.7817 | 0.8973 |

### 3.4 Results and analysis

To validate the performance of the proposed AEP-Net, we show the error prediction results of the different kinds of auto-generated masks in IBSR v2.0 dataset in Table 1. We only show the overall error prediction performance of the ACDC dataset due to the space. In IBSR v2.0 dataset, the masks are divided into five groups in terms of which $DSC$ range the $Seg.DSC$ belongs to. We also list the number of masks in the second column, and the last two rows show the overall performance on all masks. It is noticed that the larger error regions in the low-quality masks are easier to be found, while the small, thin error regions near the class boundaries in the good quality masks are harder to be detected in the absence of ground truths [5]. The proposed AEP-Net performs well in both five groups. Specifically, it can be observed in the fifth row that even the masks with good quality, our proposed AEP-Net still performs well for detecting the small, thin error regions. We achieve a DSC of 0.8358 on the IBSR v2.0 dataset and a DSC of 0.8164 on the ACDC dataset. More experimental results of the ACDC dataset will be shown in supplementary materials.

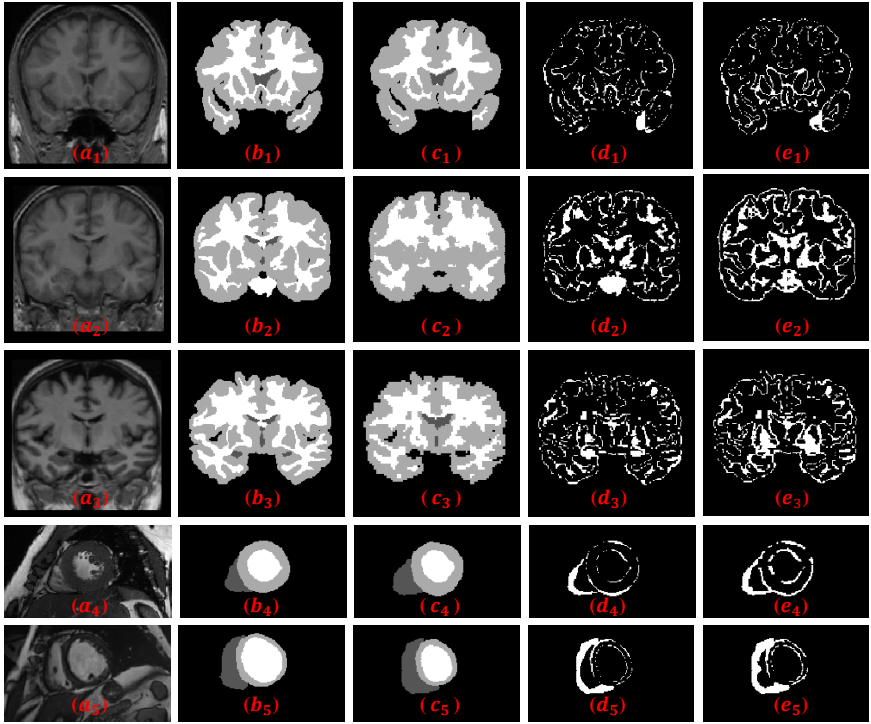

**Fig. 4.** The error prediction results on IBSR v2.0 dataset and ACDC dataset, (a) Original images (b) Ground truths (c) Generated masks (d) Real error maps (e) Predicted error maps of the generated masks

**Table 2.** Ablation experiments

| Method | Mask type | Num | DSC | Acc | Prec | Recl |
|---|---|---|---|---|---|---|
| 3D U-Net | IBSR v2.0 | 234 | 0.7823 | 0.9612 | 0.7925 | 0.8205 |
| AEP-Net no CEU | IBSR v2.0 | 234 | 0.8153 | 0.9785 | 0.8054 | 0.8450 |
| AEP-Net | IBSR v2.0 | 234 | **0.8358** | **0.9882** | **0.8190** | **0.8658** |

**Ablation study.** In order to prove the effectiveness of the network design. We compare the AEP-Net with a traditional 3D U-Net structure, which has the similar number of parameters as the AEP-Net, and it takes the concatenation of the original images and the segmentation masks as input. We also remove the CEU from the AEP-Net for ablation experiments. All comparative experiments use the same data division. As shown in Table 2, the higher average $DSC$ produced by the AEP-Net illustrates the framework design is more suitable for the error prediction task.

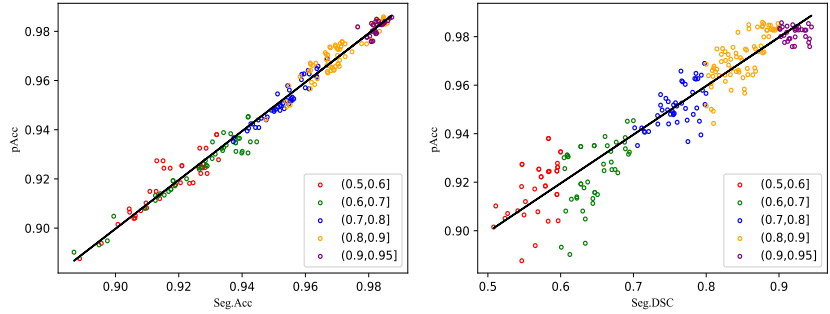

**Fig. 5.** Scatter plots of (a) Seg.Acc and pAcc, and (b) Seg.DSC and pAcc. The different colors represent different Seg.DSC range.

**Segmentation Quality Assessment.** To evaluate the effectiveness of the segmentation quality assessment inferred from the predicted error map, we define the $pAcc$ as the percentage of the correctly classified voxels to the total voxels inferred from the predicted error map, which denotes the predicted segmentation accuracy. Based on the error prediction results of the IBSR v2.0 dataset, we calculate the Pearson correlation coefficient ($PCC_a$) and the mean absolute error ($MAE$) between the $Seg.Acc$ and the $pAcc$, and the $PCC_d$ between the $Seg.DSC$ and the $pAcc$. We also draw scatter plots of these two groups of metrics. The high $PCC_a$ and $PCC_d$, low $MAE$ shows the segmentation quality can be accurately assessed according to the predicted error map.

$$PCC_a = 0.9873, PCC_d = 0.9308, MAE = 0.0031 \tag{5}$$

## 4 Conclusion

To address the need for fine-grain segmentation quality assessment of per case in image-based automated medical diagnosis system or medical research, we propose a collaborative boundary-aware context encoding framework for accurate error prediction in medical image segmentation. The proposed framework shows good performance on the auto-generated masks of different quality, especially for detecting the small and thin error regions, which ensures that the quality measurement derived from the predicted error map reflects the segmentation quality accurately and has a high reference value. In the future, we will study multi-class error map prediction and how to improve the segmentation accuracy by using error map prediction adaptatively.

# Supplementary materials

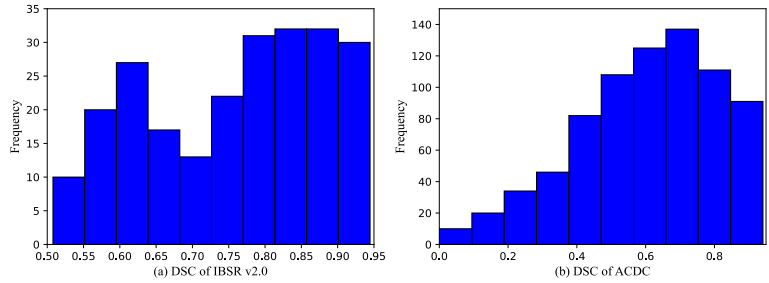

**Fig. 1.** The DSC distribution histograms of the auto-generated masks for two datasets

**Table 1.** Error prediction results of the ACDC dataset

| Mask type | Num | Seg.DSC | Seg.Acc | DSC | Acc | Prec | Recl |
|---|---|---|---|---|---|---|---|
| (0,0.2]   | 38  | 0.1267 | 0.7491 | 0.9215 | 0.9754 | 0.9568 | 0.8971 |
| (0.2,0.4] | 120 | 0.3158 | 0.8045 | 0.9171 | 0.9866 | 0.9166 | 0.9222 |
| (0.4,0.6] | 261 | 0.5071 | 0.8427 | 0.8622 | 0.9888 | 0.8379 | 0.9160 |
| (0.6,0.8] | 239 | 0.6872 | 0.9097 | 0.8277 | 0.9895 | 0.7972 | 0.8908 |
| (0.8,1]   | 142 | 0.8683 | 0.9873 | 0.6000 | 0.9887 | 0.5231 | 0.8532 |
| ACDC      | 800 | 0.5782 | 0.8782 | 0.8164 | 0.9880 | 0.7817 | 0.8973 |

**Table 2.** Ablation study on ACDC dataset

| Method | Mask type | Num | DSC | Acc | Prec | Recl |
|---|---|---|---|---|---|---|
| 3D U-Net       | ACDC | 800 | 0.7723 | 0.9654 | 0.7542 | 0.8410 |
| AEP-Net no CEU | ACDC | 800 | 0.7932 | 0.9785 | 0.7654 | 0.8550 |
| AEP-Net        | ACDC | 800 | **0.8164** | **0.9880** | **0.7817** | **0.8973** |

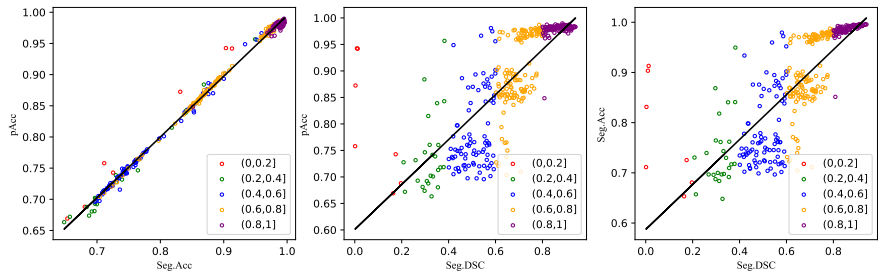

**Fig. 2.** Scatter plots of (a) Seg.Acc and pAcc, and (b) Seg.DSC and pAcc (c)Seg.DSC and Seg.Acc. The different colors represent different Seg.DSC range.

As shown in Fig. 2, we randomly selected 400 masks from ACDC dataset to draw scatter plots. The $Seg.DSC$ sometimes fails to reflect the overall accuracy of per case due to the small target region and the wide range of the in-plane dimension. However, it can be observed from (b) and (c) that the distribution of two scatters is very similar, which illustrates the segmentation accuracy inferred from the error map still performs well. Additionally, we calculate the Pearson correlation coefficient ($PCC_a$) and the mean absolute error ($MAE$) between the $Seg.Acc$ and $pAcc$, and the $PCC_d$ between the $Seg.DSC$ and $pAcc$. The high $PCC$ and low $MAE$ also verify the efficacy of our proposed AEP-Net.

$$PCC_a = 0.9979 \ \ MAE = 0.0042 \ \ PCC_d = 0.8012$$